**Ecologically mapped neuronal identity: Towards standardizing activity across heterogeneous experiments**

Kevin Luxem[1] and David Eriksson[2*]

1. Department of Cellular Neuroscience, Leibniz Institute for Neurobiology, Brenneckestr. 6, 39118, Magdeburg, Germany

2. Institute of Physiology, University of Freiburg, Hermann-Herder-Str. 7, 79104 Freiburg, Germany.

*(david.eriksson@physiologie.uni-freiburg.de).

**Abstract**

The brain's diversity of neurons enables a rich behavioral repertoire and flexible adaptation to new situations. Assuming that the ecological pressure has optimized this neuronal variety, we propose exploiting naïve behavior to map the neuronal identity. Here we investigate the feasibility of identifying neurons "ecologically" using their activation for natural behavioral and environmental parameters. Such a neuronal ECO-marker might give a finer granularity than possible with genetic or molecular markers, thereby facilitating the comparison of the functional characteristics of individual neurons across animals. In contrast to a potential mapping using artificial stimuli and trained behavior which have an unlimited parameter space, an ecological mapping is experimentally feasible since it is bounded by the ecology. Home-cage environment is an excellent basis for this ECO-mapping covering an extensive behavioral repertoire and since home-cage behavior is similar across laboratories. We review the possibility of adding area-specific environmental enrichment and automatized behavioral tasks to identify neurons in specific brain areas. In this work, we focus on the visual cortex, motor cortex, prefrontal cortex, and hippocampus. Fundamental to achieving this identification is to take advantage of state-of-the-art behavioral tracking, sensory stimulation protocols, and the plethora of creative behavioral solutions for rodents. We find that motor areas might be easiest to address, followed by prefrontal, hippocampal, and visual areas. The possibility of acquiring a near-complete ecological identification with minimal animal handling, minimal constraints on the main experiment, and data compatibility across laboratories might outweigh the necessity of implanting electrodes or imaging devices.

**Significance Statement**

All laboratories may have recorded the activity in 75 million neurons within 5 to 10 years, corresponding to a complete coverage of the mouse brain. To accumulate this emerging data-set in a reusable way, neuronal units need to have a generalized identity. Here we propose a functional identity bounded by the animal's ecology. The ECO-mapping may not only facilitate collaboration within and across laboratories, but it may also aid the interpretation of brain manipulations.

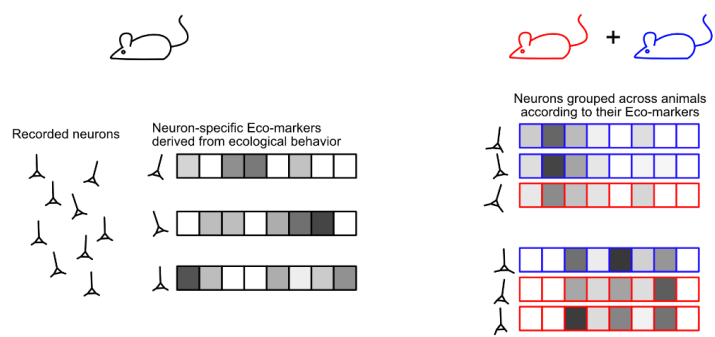

## Introduction

With the rapid development of imaging and extracellular recording techniques, all laboratories may have recorded 75 million neurons within 5 to 10 years, corresponding to a complete coverage of the mouse brain. To fully harvest this emerging data, how can we furnish the accumulation of data across institutions while retaining the creativity and individuality of each laboratory?

In one of the most significant collaborative efforts, the "International Brain Laboratory", a behavioral decision task for mice was standardized across multiple labs. The resulting initial and trained behavior was comparable across all participating laboratories (The International Brain Laboratory et al. 2021) and later extended to large-scale extracellular recordings (Laboratory et al. 2022). Such an effort allows the divide- and conquer of the mouse decision process across different areas, circuits, and laboratories. Another effort focused on standardizing high-quality extracellular recordings across many cortical and subcortical visual areas in the mouse (Siegle et al. 2021). By rigorously applying many quality criteria, one could minimize the risk that a certain neuronal response was due to suboptimal experimental procedures.

Although the above projects cover many behavioral and stimulus parameters, they do not support a bi-directional data exchange beyond the respective institutions and their carefully designed behavioral task. If the goal is accumulating and exchanging neuronal data independent of the behavioral task, we need a common ground for identifying neurons.

In state-of-the-art experiments, neurons have been identified *in-vivo* based on their genetics, morphology, and connectivity. Although this is partly feasible using imaging (Kerlin et al. 2010; Khan et al. 2018; Okun et al. 2015), it is not so for extracellular recordings. Furthermore, neuronal morphology, genetics, and connectivity might not aid species comparison. For example, the mammalian pyramidal neurons evolved single apical dendritic trees (DeFelipe and Fariñas 1992), whereas birds evolved multipolar dendritic trees (Spool et al. 2021; Reiner, Stern, and Wilson 2001; Y. Wang, Brzozowska-Prechtl, and Karten 2010; Ahumada-Galleguillos et al. 2015). Despite this, mammals and birds can solve similar tasks (Niu et al. 2022). Thus, ecology may give a bridge between species through which we can appreciate the significance of differences in, for example, morphology.

The importance of ecology for defining neuronal types has received support from studies showing that the mere behavioral and stimulus tuning of neurons is related to the genetic markers of the cell (Adesnik et al. 2012; Miao et al. 2017; Kepecs and Fishell 2014; Vlasits, Euler, and Franke 2019). Moreover, the statistics of neuronal activity patterns improve the genetic classification (Bugeon et al. 2021). Overall, this points towards identifying neurons using neuronal activity and animal ecology.

How would a first principle ecological identification be performed? It appears reasonable to consider that neuronal diversity governs the behavioral repertoire (**Fig. 1A**). Here, we assume that evolution and development optimize this behavior (Gomez-Marin and Ghazanfar 2019). Accordingly, it would be suboptimal to identify neurons using 1) artificial tasks that are 2) limited in scope (**Fig. 1B-C**). A natural task would allow the screening of more contexts, and if we assume that the brain optimizes neuronal variety for the ecology of the animal, the identity will be more meaningful.

With a complex ecological environment, we maximize the chance for differentiating between any two neurons, which is impossible with genetic markers given that hundreds of neurons can have the same genetic marker. Each neuron will have a unique connectivity with certain brain areas thereby modulating neuronal tuning in a context specific way (Carmena et al. 2003; Maier, Logothetis, and Leopold 2007; Kumano, Suda, and Uka 2016; Nigam, Pojoga, and Dragoi 2021). Here, we suggest exploiting this dynamic tuning to increase the discriminability between different neurons.

Since we indirectly take advantage of varying contexts, this puts explicit experimental constraints on the behavior and environment. Different mapping periods must cover the same types of behaviors. Furthermore, to equilibrate representational drift (Deitch, Rubin, and Ziv 2021) across mappings, the behavioral/environmental statistics preceding the mapping periods should be similar. Although it is impossible to satisfy those stationarity constraints perfectly, it is conceivable that the behavior in a home-cage environment is one of the best candidates. Here constraint 1 is satisfied since each animal engages in daily routines such as running, exploring, grooming, and eating. Constraint 2 is satisfied since the pre-mapping behavior can be standardized, for example, by the time of the mapping. Finally, an ecological setting imposes a minimal challenge for the brain, thereby minimizing representational drift (Driscoll, Duncker, and Harvey 2022).

Home-cage-like behavior can cover a large part of natural behavior (Weissbrod et al. 2013) and willl be similar across laboratories (Grieco et al. 2021; Voikar and Gaburro 2020; Robinson, Spruijt, and Riedel 2018). Moreover, it is less prone to animal-handling differences across laboratories, offering an excellent standardization basis. Here we propose that homepage behavior can cover a large portion of neuronal variety (**Fig. 1D**). This variety can be further extended by adding brain area-specific environmental enrichment (**Fig. 1E-F**).

We focus on home-cage compatible tasks for identifying neurons in the visual cortex, motor cortex, prefrontal cortex, and hippocampus of mice and rats. Overall, we hope that the area-specific environmental enrichment reviewed here can serve as a starting point for developing ecologically-based fine-grained neuronal identification. Such identification could be both generic and pragmatic. Generic because animal behavior and environment are common across individuals and many species. Pragmatic because natural behavior lends itself to minimal experimental interference and automatic procedures.

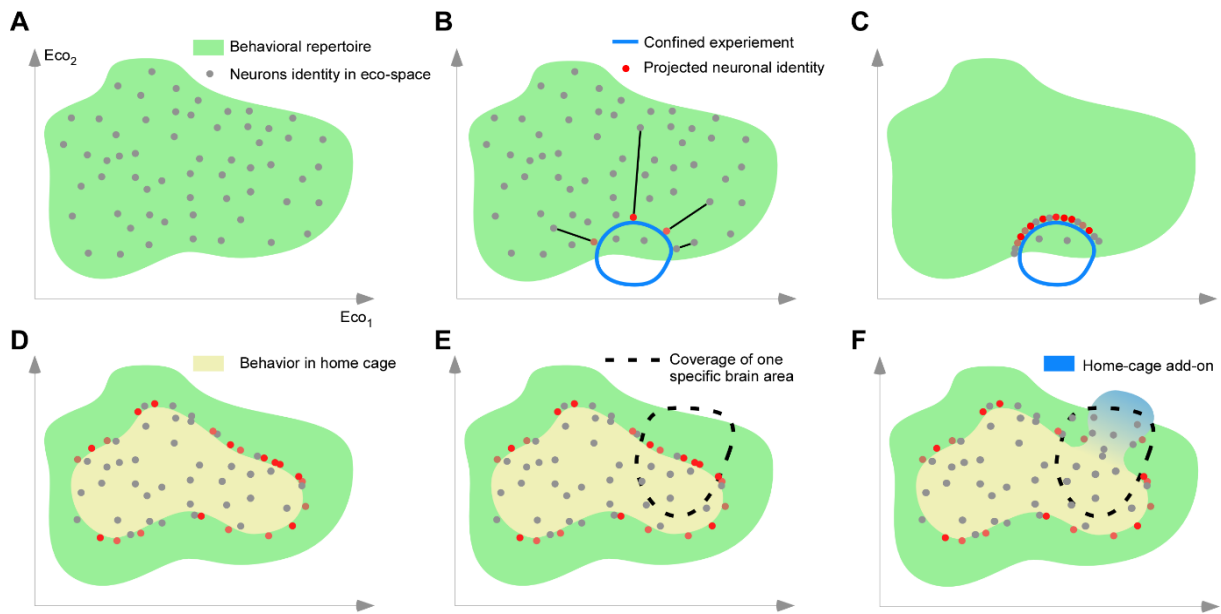

**Figure 1.** Ecological neuronal identity depends on the type of behavior and environment. A: For illustration, the identity of the neurons (grey dots) is defined by the location in the two-dimensional ECO-space. Each dot's location quantifies that corresponding neuron's activity-tuning for two different behavioral and environmental parameters: $ECO_1$ and $ECO_2$. ECO dimensions can, for example, be rewarded firing rate modulation, running firing rate modulation, etc. B-C: Non-ecological identity using a confined experiment, e.g. a classical experiment in which a specific set of scientific hypotheses is addressed. Neuronal identities seen for a rich behavioral repertoire are projected and clipped by the limited dynamic range of the confined experiment. Neurons with identities within the dynamic range of the experiment (within the blue circle) will remain undistorted. D: Home cage behavior can capture a significant fraction of the neuronal variety. E: For neurons in a given brain area, the home cage behavior covers a subset of the neuronal variety. F: An area-specific home cage add-on allows the correct identification of more neurons in a given brain area.

**Results**

Here we explore the feasibility of extracting neuronal identities from ecologically appropriate behavior and environments. To this end, multiple values define the identity of a neuron. Each value quantifies how a neuron responds to a particular behavioral and environmental parameter (ECO-dimension). Over the last 70 years, researchers have systematically investigated how neurons in various areas co-activate with different motor and stimulus protocols. Here we review commonly used tuning variables to classify neurons in a selected set of brain areas: visual-, motor-, hippocampal-, and frontal areas. We will refer to identifying neurons using behavioral and environmental variables as ECO-mapping.

We will assume that freely moving recordings (imaging or electrophysiology), as opposed to head-fixation, are necessary for the ECO-mapping in most brain areas. For example, rearing-specific neuronal activations are common in diverse areas such as the motor cortex, hippocampus (Lever, Burton, and O'Keefe 2006; Deacon, Croucher, and Rawlins 2002; Harley and Martin 1999), and prefrontal cortex (Gemmell, Anderson, and O'Mara 2002). Although head fixation is more suitable for sensory areas for practical/experimental reasons, recent work shows that head direction modulates the activity of individual neurons in the primary visual cortex (Parker et al. 2022). Head alignment using sensors and reward delivery (Eriksson et al. 2021) might be an alternative to head fixation for more controlled sensory stimulation in the freely moving animal. Furthermore, the constant improvement of head-held multi-photon microscopes makes them more lightweight and suitable for natural behavior (Zong et al. 2022). Finally, although head fixation facilitates virtual reality on a floating ball, there are powerful solutions for virtual reality in freely moving animals (Kaupert et al. 2017).

Animal behavior research has opened up a wonderland for freely moving animals with creative solutions such as a joystick in the home cage (Bollu et al. 2019), touchscreens for animals (Beraldo et al. 2019), Wisconsin card sorting test for rodents (Birrell and Brown 2000), virtual reality for freely moving animals (Kaupert et al. 2017), and automatic exchange of animals between home cages (Kaupert et al. 2017). Certain combinations of those solutions may cover more exotic ecological dimensions of different brain areas, allowing us to test neuronal activity tuning in multiple contexts. Here we will investigate how to structure this combination for a particular brain area in a home-cage environment. To this end, for each brain area, we divide the neuronal identification into three task-complexity grades dependent on the amount of effort required by the experimenter and the animal:

**Grade 1**: Low effort. Naïve/everyday behavior without learning, no cost-function (e.g., passive sensory viewing or locomotion).

**Grade 2**: Moderate effort. Naïve behavior with learning, intuitive cost-function (e.g., reaching for food or water maze). The animal learns with a static/non-changing task.

**Grade 3**: High effort. Abstract learning/cost function (e.g., grabbing and moving a joystick or delays). The behavior task has to be gradually modified during training (for example, gradually increasing the delay between stimulus and behavioral response) to ensure that the animal learns the task.

Since **grade 3** tasks may reduce the performance of the main experiment, the goal is to use as many **grade-1** and **2** tasks and, if possible, to integrate **grade 3** tasks in the main scientific experiment. At the same time as **grade 3** requires extensive training, it might be the least essential grade for defining neuronal identities. Since evolution and development cannot predict future situations, it is conceivable that animals shapes existing **grade 1** and **2**-governing circuits when learning a complex **grade 3** task (Sadtler et al. 2014). The more varied its naïve behavior is, the more likely it may be that a circuit can be reshaped (Gomez-Marin and Ghazanfar 2019). Therefore, the "enabler" of flexible behavior is the animals' naïve toolbox of neuronal diversity. Thus, **grade 1** and **2** ECO-mapping might be sufficient for explaining why individuals solve and perform tasks differently (The International Brain Laboratory et al. 2021).

Identification of general behavior

Behavior can be measured in different granularities, from the center-of-mass location of the animal to tracking all body parts with markerless pose estimation (Pereira et al., 2020). To accurately map behavioral specific neuronal subsets and ensure the robustness of the ECO-mapping approach, it is crucial to quantify and segment behavior on a fine-grained scale. This requires measuring the kinematics of the animal with markerless pose estimation (Luxem et al., 2023) and identifying behavioral patterns or motifs in the dynamic changes of the animal's pose. VAME (Variational Animal Motion Embedding), a recently proposed algorithm, excels in this task as it not only identifies the structure of complex pose dynamics but also embeds it into a lower-dimensional space to identify ecologically relevant behaviors in a self-supervised fashion (Luxem et al., 2022). The algorithm's lower-dimensional vector representation of behavior enables easy correlation with any embedding of the neuronal activity representation. Furthermore, using virtual marker positions in the home-cage setup makes the approach highly reproducible between laboratories, rendering VAME a particularly powerful tool for identifying ECO-mapping relevant behaviors.

Visual system

Examples of neuronal tunings that can be tested with **grade 1** tasks are temporal frequency- or visual orientation tuning, passive attention to sudden/novel stimulus, and tuning to behavioral variables such as eye movements and locomotory modulation (for the head-fixed animal see (Siegle et al. 2021)). Attentional modulation will be regarded as **grade 2** if there is no delay and **grade 3** if there is a delay between stimulus and behavioral response. The possibility of studying selective visual attention in the head-fixed mouse is relatively new (L. Wang and Krauzlis 2018). In a spatial selective attention task, mice do not move their eyes (Kanamori and Mrsic-Flogel 2022), suggesting covert attention in line with the lack of a fovea but see (van Beest et al. 2021). The head fixation even allows measuring the membrane potential during an attention paradigm (Speed et al. 2020).

For freely moving animals, one can use the 5-choice serial reaction time task (Carli et al. 1983; Robbins 2002; Bari, Dalley, and Robbins 2008). This task addresses many attentional aspects appropriate for higher visual areas, such as reaction time, accuracy, % omissions, incorrect errors, etc. (Young et al. 2009). Here, head tracking (You and Mysore 2020) and head-mounted eye-tracking (Holmgren et al. 2021) are necessary controls. As head- and eye-tracking technology becomes more mature, it is feasible to do complex attentional tasks using touchscreens (Wicks et al. 2017; S. Li et al. 2021). Given that neuronal tuning and visual processing changes in the freely moving animal, a future goal must be to optimize head and eye-tracking in the freely moving animal.

Motor system

In the motor system, **grade 1** behaviors are spontaneous locomotion and exploratory behaviors, emergent in a home cage environment without training (Mimica et al. 2018). Body tracking tools allow the identification of dozens of factors, such as paw, head, and tail positions, rearing, exploratory, and turning behavior (Dunn et al. 2021; Mathis et al. 2018). **Grade 2**: Reaching behaviors and ladder walking are natural movements that require fine-tuning training (Metz and Whishaw 2009; Churchland et al. 2012; Ames, Ryu, and Shenoy 2019). Those behaviors give information about neuronal tuning for a successful movement, failed movement, reward, etc. **Grade 3**: For example, grabbing something (e.g., a joystick) (Wagner et al. 2020) and pushing it away from the body is unnatural since if an animal grabs something, its purpose is most likely nutrition. Delayed movement tasks address planning neurons (Wagner et al. 2020). An alternative to delayed movements is to use minimally repetitive movements, which allows the separation of planning, motor, and sensory ECO-dimensions and their neuronal activity markers during freely moving exploratory (**Grade 1**) as well as for haptic manipulation (**Grade 2** and **3**) behaviors (Eriksson et al. 2021). Haptic manipulation devices ultimately address proprioception and force tuning. Therefore, the powerful combination of haptic joystick devices and multi-camera video tracking invites ecological mapping in the motor system.

Hippocampus

For the hippocampus, a **grade 1** task is everyday behavior when the animal explores its home cage. To identify neurons based on their tuning, it becomes important to separate between different aspects such as space, time, duration, feature, and reward (O'Keefe and Krupic 2021; Eichenbaum et al. 1999). Therefore, the same object feature may appear at different places (Inayat et al. 2021), or different features to appear at the same place. The object-position task is formalized with the "Novel object recognition task" (NORT) and requires no training (Clark, Zola, and Squire 2000; Ainge et al. 2006). **Grade 2**: The animal will fine-tune its natural behavior repertoire, for example, to find a platform in the Morris water maze (Morris et al. 1982). Such a task will provide classification in terms of reward and learning. **Grade 3**: For reverberatory tuning, one can introduce a delay to the NORT (Hammond, Tull, and Stackman 2004) or use a working memory task such as the radial arm maze task (Winters et al. 2004). Both of those tasks require the animal to hold a strategy in memory. Thus, those tasks address neuronal identities according to memory formation, sustaining, and execution.

Physical objects rather than virtual (visual) allow the combination of ecologically important modalities such as texture, odor, shape, reward, and even auditory and temperature. Given the importance of objects, position, and time in the hippocampus, it is surprising that there is no way to replace physical objects in a home-cage environment automatically. To this end, multiple object carousels are possible (Prigg et al. 2002).

Rodent prefrontal cortex

For the rodent prefrontal cortex (rodent PFC) (Laubach et al. 2018), **grade 1** tasks are social behavior, reward prediction and error (Starkweather, Gershman, and Uchida 2018), and exploration. Example situations for social behavior are animal versus object, novel versus known animal (Levy et al. 2019), social play (van Kerkhof et al. 2013), and empathy (C.-L. Li et al. 2018). Behaviors related to exploration are search direction or posture (Mimica et al. 2018). For the prefrontal cortex, a **grade 2** task would cover working memory, attention, impulsivity, and rule-based. Those aspects can be covered with a variant of the Wisconsin card sorting task for rodents, the attentional set-shifting task (ASST) (Birrell and Brown 2000). According to a **grade 2** task, the ASST cost function is intuitive, and a rat can learn each step in the task without each step being gradually changed. According to a **grade 3** task, the complexity of the task gradually increases across those steps. To address neuronal tuning for endogenous cognitive variables, one may use model-derived single-trial cognitive rule-based (Schmitt et al. 2017), and evidence accumulation (Piet et al. 2017) quantities.

It is possible to cover many ecological dimensions relevant to rodent PFC, such as social, exploration, reward, working memory, impulsivity, rule-based, attention, and motor dimensions. Nevertheless, it will be hard to cover those aspects completely since each aspect might depend on multiple sensory modalities, but see a recent study for modality invariance (Rikhye, Gilra, and Halassa 2018). Thus, a remaining question is whether it is technically possible to test multiple modalities in the home-cage environment. Is it possible and meaningful to extend the modalities of the ASST (texture, medium, and odor) with auditory and visual cues? Related questions are which sensory modalities are ecologically meaningful, under which circumstances/contexts are they meaningful, and how do different modalities correlate? According to the philosophy of ECO-mapping, two ecologically-correlated modalities do not need to be (artificially) decorrelated experimentally, thereby saving experimental complexity.

**Discussion**

Research in behavioral ecology examines how external factors can influence species to evolve and develop a certain range of behaviors. Here we take this notion further by assuming that the ecological pressure dictates the neuronal diversity/variety. According to this view, ecological markers are the optimal way to identify neurons. Behavioral and environmental factors are more tangible for in-vivo experiments than genetic, neuronal morphology, or connectivity factors. Here we investigate the feasibility of this approach in four brain areas. We reviewed which stimulus and behavioral variables are suitable for identifying neurons for each brain region using state-of-the-art experimental designs. We find that a large neuronal identification space can be covered using relatively simple, home cage-like behavior and area-specific environmental enrichment. Such an approach will allow us to identify neurons using a larger range of behavioral and environmental aspects, and the ecological perspective will make this identification more meaningful. This standardized eco-mapping and the experimental freedom of the main experiment come at the cost of requiring chronic recordings and automatized home-cage recording facility.

Limitations of the ECO-mapping

Given that we can record in an area-specifically enriched home cage, what are the fundamental limitations of this ECO-mapping? Strictly speaking, the behavior should be stationary across mappings. Natural behavior for which the brain is optimized, for example, eating, drinking, exploring, and grooming, are all executed daily, resulting in behavioral statistics that are approximately stationary across days. Moreover, since we assume that the brain is minimally challenged during home-cage behavior we hypothesize that plastic processes are at a minimum, thereby conserving circuit function. Nevertheless, the brain might constantly fine-tune properties even in a perfectly stationary environment. For example, the bouton turnover is around 6% per week many brain areas in adult un-trained home-cage housed mice (De Paola et al. 2006; Marik et al. 2010).

Which changes in neuronal tuning would such a bouton dynamics cause? Although no study have quantified the representational drift (Driscoll, Duncker, and Harvey 2022) for a home-cage environment, many studies have addressed this in various learning-driven artificial settings finding a change in representation after days to weeks (Schoonover et al. 2021; Driscoll et al. 2017; Ziv et al. 2013; Kentros et al. 2004; Marks and Goard 2021; Rubin et al. 2015). Here it is interesting to note that the hippocampus and olfactory cortex show a relatively substantial change in representation after one week (Ziv et al. 2013; Schoonover et al. 2021). Curiously, place field tuning can disappear and reappear with a minor change in the place field location (Ziv et al. 2013). This hippocampal extreme illustrates that neurons can change their tuning rapidly and dynamically depending on the context while preserving the distribution across neurons, reflecting that synapses are relatively stable across days, thereby retaining an approximately stable population code. Thus in the extreme case of the hippocampus, the ECO-mapping may enable us to sample the distribution across the neuronal population. The exact timing and context of tuning changes remain to be studied. Moreover, future studies may test the representational drift for more natural behaviors that are not learning intensive. To understand this, "Great efforts must be taken to record and model as many features of behavior as possible to better understand the evolving relationship between neuronal activity and behavior" (Driscoll, Duncker, and Harvey 2022).

Ultimately, ECO-mapping should allow comparison across animals and even across species. Classically, neuronal responses are related across animals using subspace methods that detect commonalities across manifolds of individual animals (Rubin et al. 2019; Pandarinath et al. 2018; Melbaum et al. 2022). Subspace methods are complex, hard to interpret from a behavioral perspective, and they only allow a coarse identification of neuronal groups. To achieve more interpretable, neuronally resolved identities, one should take advantage of the fact that each neuron represents the environment and behavior uniquely. Here we assume that units with similar ECO-markers will be comparable and can be "pooled" across animals. To what extent will the neuronal distribution (of ECO-markers) in a given brain region be similar for two different animals? Can two animals have non-overlapping, modular neuronal types, or is there an overlap across animals with mere quantitative differences? Rather than each animal having a unique non-overlapping set of cells, recent evidence from stellate cells in the entorhinal cortex suggests that cellular properties vary along a continuum where the major difference between animals is the average "set-point" (Pastoll et al. 2020). Even across mammalian species, differences in neuronal firing regimes are dictated by processed information rather than the species per se (Mochizuki et al. 2016). Nevertheless, it is well known that animals have a personality and that different animals solve behavioral tasks in different ways. An open question is if this personality is due to region- and animal-specific set-points or if the neurons are functionally qualitatively different across animals. Animal differences in set points would be compatible with the ECO-mapping, whereas qualitatively different neurons would be harder to reconcile with the ECO-mapping approach.

ECO-mapping for controlled causal manipulations

Neuronal activity is vital for guiding experimental brain manipulations. To this end, areas with strong task-specific activity are suitable targets for loss-of-function manipulations. The relation between the suppressed area's neuronal activity and downstream areas' suppression effects has a general functional form (i.e., impact function) for many brain mechanisms (Eriksson 2017; Eriksson, Ramanathan, and Veit 2022). Optimally the activity of all neurons in the targeted area should be considered for guiding and interpreting manipulations. However, current recording techniques typically only cover a subset of the neurons in a given area, causing quantifiable statistical biases (Levina, Priesemann, and Zierenberg 2022).

One way to address the subsampling problem is to pool neurons from multiple animals. Given that it is common practice to ascertain the robustness of an experimental result using recordings from multiple animals, the implicit assumption is that different animals have similar distributions of neuronal types. Statistically, the unrecorded neurons in one animal should have their counterparts in the recorded neurons of other animals. Thus, it ought to be possible to populate a recording session with virtual neurons -based on neurons recorded in another animal. How does one predict the responses of neuron type X (recorded in animal and session A) when it is in a different context of animal and session B? The eco-mapping powerfully addresses this issue since the ECO-markers quantify how a neuron responds to a wide range of ecologically relevant behavioral/environmental variables.

Can the ECO-mapping be used to predict the activity of un-recorded units on a single trial level? In a classical study, the activity of neurons with a particular orientation preference in the visual cortex is strongly correlated with the instantaneously encoded orientation of surrounding neurons (Tsodyks et al. 1999). This and similar results (Kenet et al. 2003, for an alternative view Stringer et al. 2019; Okun et al. 2012) can be interpreted as the correlation between the ongoing activity in neurons that resembling the visual statistics of the environment (Geisler 2008). This suggests that the correlation between neurons can be exploited to predict unrecorded neurons' single-trial activity. Another study showed that a neuron's choristers- and soloists-identity dictated how its instantaneous activity was correlated with the activity of surrounding neurons (Okun et al. 2015). Finally, endogenous, single-trial relations between neurons predict the behavioral/environmental variables and, importantly, the dimension-reduced single-trial dynamics of unrecorded neurons: neurons recorded in another animal (Rubin et al. 2019; Pandarinath et al. 2018; Melbaum et al. 2022). Those studies suggest that the population activity conveys information for predicting the single-trial activity of un-recorded neurons, especially if neurons' behavioral and environmental identity is known.

Conclusion

It would be wonderful if Neuroscience were like physics, for which a global statistical- or deterministic phenomena could be condensed to a mean-field description. Instead, Neuroscience is the study of a system in which the particular environment of that very organism shapes every neuron and every synapse. This detailed adjustment to the many aspects of an environment allows the organism to be flexible. Therefore, any behavior is of interest. The ecological standardization of neural activity suggested here may facilitate collaboration between laboratories independently of the behavior studied. Such standardization may allow us to understand the differences in personalities across animals and why different animals apply different strategies for the same task.